\documentclass[ALICE,manyauthors]{cernphprep}
\usepackage[comma,square,numbers,sort&compress]{natbib}
\usepackage{hyperref}
\usepackage{lineno}

\newcommand{\sqrts}{\sqrt{s_{\scriptscriptstyle \rm NN}}}
\newcommand{\sqrtsNN}{\sqrt{s_{\scriptscriptstyle \rm NN}}}

\newcommand{\av}[1]{\left\langle #1 \right\rangle}

\newcommand{\gev}{\mathrm{GeV}}

\newcommand{\tev}{\mathrm{TeV}}

\newcommand{\PbPb}{\mbox{Pb--Pb}}
\newcommand{\pt}{p_{\rm T}}

\newcommand{\DtoKpi}{{\rm D}^0 \to {\rm K}^-\pi^+}
\newcommand{\DtoKpipi}{{\rm D}^+\to {\rm K}^-\pi^+\pi^+}
\newcommand{\DstartoDpi}{{\rm D}^{*+} \to {\rm D}^0 \pi^+}

\newcommand{\Dzero}{{\rm D^0}}

\newcommand{\Dstar}{{\rm D^{*+}}}

\newcommand{\Dplus}{{\rm D^+}}

\newcommand{\Jpsi}{{\rm J}/\psi}

\newcommand{\dEdx}{{\rm d}E/{\rm d}x}

\renewcommand{\PbPb}{\mbox{Pb--Pb}}
\newcommand{\Npart}{N_{\rm part}}

\newcommand{\RAA}{R_{\rm AA}}
\newcommand{\TAA}{T_{\rm AA}}

\begin{document}

\begin{titlepage}

\PHyear{2015}
\PHnumber{151}                 
\PHdate{17 June}              

\title{Centrality dependence of high-$\mathbf{\textit{p}}_{\rm T}$ D meson suppression\\ in $\PbPb$ collisions at 
$\mathbf{\sqrt{\textit{s}_{\mathbf \rm NN}}}$ = 2.76 TeV}
\Collaboration{ALICE Collaboration%
         \thanks{See Appendix~\ref{app:collab} for the list of collaboration 
                      members}}
                     
\ShortAuthor{ALICE Collaboration}
\ShortTitle{Centrality dependence of high-$\pt$ ${\mathbf{\rm D}}$ meson suppression}

\begin{abstract} 

The nuclear modification factor, $\RAA$, of the prompt charmed mesons $\Dzero$, $\Dplus$ and $\Dstar$, and their antiparticles,
was measured with the ALICE detector in $\PbPb$ collisions at a centre-of-mass energy $\sqrts = 2.76~\tev$ in two transverse momentum intervals, 
$5<\pt<8~\gev/c$ and $8<\pt<16~\gev/c$,
and in six collision centrality classes. 
The $\RAA$ shows a maximum suppression of a factor of  5--6 in the 10$\%$ most central collisions.  
The suppression and its centrality dependence are compatible within uncertainties with those of charged pions. 
A comparison with the 
$\RAA$ of non-prompt ${\rm J}/\psi$ from B meson decays, measured by the CMS
Collaboration, hints at a larger suppression of D mesons in the most central collisions.

\end{abstract}


\end{titlepage}
\setcounter{page}{2}

\section{Introduction}
\label{sec:intro}
When heavy nuclei collide at high energy, a state of strongly-interacting matter with high energy density is expected to form. 
According to Quantum Chromodynamics (QCD) calculations on the lattice, this state of matter, the so-called Quark-Gluon Plasma (QGP) is characterised by the deconfinement of the colour charge (see e.g.~\cite{karsch,Borsanyi,Borsanyi2,Bazavov}).
High-momentum partons, produced at the early stage of the nuclear collision, lose energy as they interact with the QGP constituents. This energy loss is expected to proceed via both inelastic (gluon radiation)~\cite{gyulassy,bdmps} and elastic (collisional) processes~\cite{thoma,thoma2,thoma3}. 

The nuclear modification factor $\RAA$ is used to characterise 
parton energy loss by comparing 
particle production yields in nucleus--nucleus collisions to a scaled proton--proton (pp) reference, that corresponds to 
a superposition of independent nucleon--nucleon collisions.
$\RAA$ is defined as
\begin{equation}
\label{eq:Raa}
R_{\rm AA}=
{1\over \av{T_{\rm AA}}} \cdot 
{{\rm d} N_{\rm AA}/{\rm d}\pt \over 
{\rm d}\sigma_{\rm pp}/{\rm d}\pt}\,,
\end{equation}
where ${\rm d}\sigma_{\rm pp}/{\rm d}\pt$ and ${\rm d} N_{\rm AA}/{\rm d}\pt$ are the transverse momentum ($\pt$) differential cross section and yield in proton--proton and nucleus--nucleus (AA) collisions, respectively. $\av{T_{\rm AA}}$ is the average nuclear overlap function, 
estimated within the Glauber model of the nucleus--nucleus collision geometry, and proportional to the average number of 
nucleon--nucleon (binary) collisions~\cite{glauber,glauber1}.
Energy loss shifts the momentum of quarks and gluons, and thus hadrons, towards lower values, leading
to a suppression of hadron yields with respect to binary scaling at $\pt$ larger than few $~\gev/c$ ($\RAA < 1$). 

Energy loss is expected to be smaller for quarks than for gluons because the colour charge factor of quarks is smaller than that of gluons~\cite{gyulassy,bdmps}.  
In the energy regime of the Large Hadron Collider (LHC), light-flavour hadrons with $\pt$ ranging from 5 to $20~\gev/c$ originate predominantly from gluon fragmentation (see e.g.~\cite{Djordjevic:2013pba}). At variance,
charmed mesons provide an experimental tag for a quark parent. Because 
of their large mass $m_{\rm c,b}$  ($m_{\rm c} \approx 1.3~\gev/c^2$, $m_{\rm b} \approx 4.5~\gev/c^2$~\cite{PDG}), heavy quarks are produced at the initial stage
of heavy-ion collisions in hard scattering processes that are characterised by 
a timescale $\Delta t < 1/(2\,m_{\rm c,b})\sim 0.1\,(0.01)~{\rm fm}/c$ for c\,(b) quarks.
This time is shorter than the formation time of
the QGP medium (a recent estimate for the LHC energy is about $0.3~{\rm fm}/c$~\cite{Liu:2012ax}). 
As discussed in Ref.~\cite{SaporeGravis}, this should be the case also for charm and beauty quarks produced in gluon splitting processes, if their transverse momentum is lower than about $50~\gev/c$.
Therefore,  the comparison of the heavy-flavour hadron $\RAA$ with that of pions 
allows the colour-charge dependence of parton energy loss to be tested. The softer fragmentation of gluons than that of charm quarks, and the observed
increase of the charged hadron $\RAA$ towards high $\pt$~\cite{Abelev:2012hxa}, 
tend to counterbalance the effect of the larger energy loss of gluons on the $\RAA$. The model predictions range from a rather moderate 
effect  $\RAA^{\pi}<\RAA^{\rm D}$~\cite{adsw,whdgzero,whdgone,whdgtwo} to an overall compensation $\RAA^{\pi}\approx\RAA^{\rm D}$
(as recently shown in~\cite{Djordjevic:2013pba}) in the $\pt$ interval from 5 to about $15~\gev/c$.

Several mass-dependent effects are expected to influence the energy loss for quarks
(see~\cite{SaporeGravis} for a recent review).
The dead-cone effect should reduce small-angle gluon radiation for quarks that have moderate energy-over-mass values, i.e.\,for c and b quarks 
with momenta up to about 10 and $30~\gev/c$, respectively~\cite{dk,asw,dg,wang,whdgzero}.
Likewise, collisional energy loss is expected to be reduced for heavier quarks, because the spatial diffusion coefficient that regulates
the momentum exchange with the medium is expected to scale as the inverse of the quark mass~\cite{rapp}.
In the $\pt$ interval up to about $20~\gev/c$, where the masses of heavy quarks are not negligible with respect to their momenta, 
essentially all models predict \mbox{$\RAA^{\rm D} < \RAA^{\rm B}$}~\cite{whdgzero,whdgone,whdgtwo,adsw,adil,vitev,Buzzatti:2012dy,He:2012xz,He:2014cla, Gossiaux:2012ya,Uphoff:2012gb,Alberico:2013bza,Cao:2013ita,Lang:2012cx}, which stems directly from the mass dependence 
of the quark--medium interaction and is only moderately affected by the different production and 
fragmentation kinematics of c and b quarks (see e.g.~\cite{CacciariInLastCallPred}).


A first comparison of light-flavour, charm and beauty hadron nuclear modification factors based on measurements 
by the ALICE and CMS Collaborations~\cite{Abelev:2012hxa,DRAA2010,CMSJpsi2010} from the 2010 LHC Pb--Pb data at a centre-of-mass energy $\sqrtsNN=2.76~\tev$ was presented in~\cite{DRAA2010}.
In this paper we present the centrality dependence of the D meson $\RAA$ in Pb--Pb collisions at the same energy,
measured with the ALICE detector~\cite{aliceJINST} using data from both 2010 and 2011 periods (integrated luminosities of about 2.2 and 21~$\rm \mu b^{-1}$, respectively). 
The focus here is on the study of the parton energy loss; therefore, the data are presented for
the high-$\pt$ interval 5--16~$\gev/c$, where the largest suppression relative to binary scaling was observed~\cite{DRAA2010}.
The results are compared with charged pions, measured by the ALICE Collaboration~\cite{ALICEhighptPIDRAA}, with 
non-prompt J/$\psi$ mesons, measured by the CMS Collaboration~\cite{CMSJpsi2010},
and with model predictions.

\section{Experimental apparatus and data sample}
\label{sec:detector}
The Pb--Pb collisions were recorded using a minimum-bias interaction trigger, based on the information of  the signal coincidence of the V0 scintillator detectors that cover the full azimuth in the pseudo-rapidity intervals $-3.7<\eta<-1.7$ and $2.8<\eta<5.1$~\cite{Vzero}.
The measurement of the summed signal amplitudes from the V0 detectors was used to sort the events in classes of collision centrality,
defined in terms of percentiles of the Pb--Pb hadronic cross section~\cite{centrality}.  
The trigger efficiency is $100\%$  for the events considered in this analysis, which correspond to the most central $80\%$ of the Pb--Pb hadronic cross section. 
An online selection based on the information of the V0 detectors was applied to increase the statistics of central collisions for the 2011 data sample.
An offline selection using the V0 and the neutron Zero-Degree Calorimeters (ZDC) was applied to remove background from interactions of the beams with residual atoms in the vacuum tube. Events with a reconstructed primary vertex outside the interval $\pm 10$~cm 
from the interaction point along the beam direction ($z$ coordinate) were removed.
The event sample used in the analysis corresponds to an integrated luminosity $L_{\rm int} = (21.3\pm 0.7)$~$\rm \mu b^{-1}$ in the 0--10\% centrality class ($16.4\times 10^6$ events) and
$(5.8\pm 0.2)$~$\rm \mu b^{-1}$  
in each of the 10--20\%, 20--30\%, 30--40\%, 40--50\% classes ($4.5\times 10^6$ events per class). In the 50--80\% class, where 2010 data were used, the analyzed event sample corresponds to $(2.2\pm 0.1)$~$\rm \mu b^{-1}$ ($5.1\times 10^6$ events).

The decays $\DtoKpi$, $\DtoKpipi$ and $\DstartoDpi$, and their charge conjugates, were reconstructed as described in~\cite{DRAA2010}
using the central barrel detectors, which are located in a solenoid that gene\-rates a 0.5~T magnetic field parallel to the beam direction. 
Charged particle tracks were reconstructed with the Time Projection Chamber (TPC)~\cite{tpc} and the Inner Tracking System (ITS), 
which consists of six cylindrical layers of silicon detectors~\cite{its}. Both detectors provide full azimuthal coverage in the interval $|\eta|<0.9$. 
$\Dzero$ and $\Dplus$ candidates were formed from pairs and triplets of tracks
with $|\eta|<0.8$, $\pt>0.4~\gev/c$, at least 70 associated space points in the TPC, and at least two hits in the
ITS, out of which one had to be in either of the two innermost layers. $\Dstar$ candidates were formed by
combining $\Dzero$ candidates with tracks with $|\eta| < 0.8$, $\pt > 0.1~\gev/c$, and at least three associated
hits in the ITS for the 10\% most central collisions (two in the other centrality classes). 
The decay tracks of the candidate D mesons were identified on the basis of their specific ionization energy deposition $\dEdx$ in the TPC and of their flight times 
to the Time Of Flight (TOF) detector, which has the same $\eta$ acceptance as the TPC. 
Particles were  identified as pions (kaons) by requiring the measured signal to be within three times the resolution ($\pm 3\,\sigma$) around the expected mean values of $\dEdx$ and time-of-flight for pions (kaons).
Only D meson candidates with rapidity $|y|<0.8$ were considered, because the acceptance decreases rapidly outside this interval.

\section{Data analysis}
\label{sec:Dmesons}
The selection of the D meson decay topology is mainly based on the displacement of the decay tracks from the primary vertex, 
and on the pointing of the reconstructed D meson momentum to the primary vertex~\cite{DRAA2010}. 
The raw yields were determined in each centrality and $\pt$ interval 
using fits to the distributions of invariant mass $M({\rm K^-\pi^+})$ and $M({\rm  K^-}\pi^+ \pi^+)$, in the case of $\Dzero$ and $\Dplus$ mesons,  and of the difference $M({\rm K^-} \pi^+ \pi^+)-M(\rm K^- \pi^+)$ for $\Dstar$ mesons.
The fit function is the sum of a Gaussian, 
for the signal, and either an exponential function ($\Dzero$ and $\Dplus$) or a power-law multiplied with an exponential function ($\Dstar$) 
to describe the background distribution~\cite{DRAA2010}.

For $\Dzero$ mesons, an additional term was included in the fit function to account for the so-called `reflections', 
i.e.\,signal candidates that are present in the invariant mass distribution also when the ${\rm (K,\pi)}$ mass hypothesis for the decay tracks is swapped.
A large fraction (about 70\%) of these reflections is rejected by the particle identification selection. 
The residual contribution was studied with Monte Carlo simulations (described later in this section). It was found that the reflections have a broad invariant mass distribution, which is well described by a sum of two Gaussians, and its integral amounts to about 30\% of the yield of the signal in the $\pt$ interval used in the analysis presented in this article. 
In order to account for the contribution of reflections in the data, a template consisting of two Gaussians was included in the fit. 
The centroids and widths, as well as the ratios of the integrals of these Gaussians to the signal integral, were fixed to the values obtained in the simulation (see~\cite{ALICEDmesonv2article} for more details).

\begin{figure}
  \centering
  \includegraphics[width=0.99\textwidth]{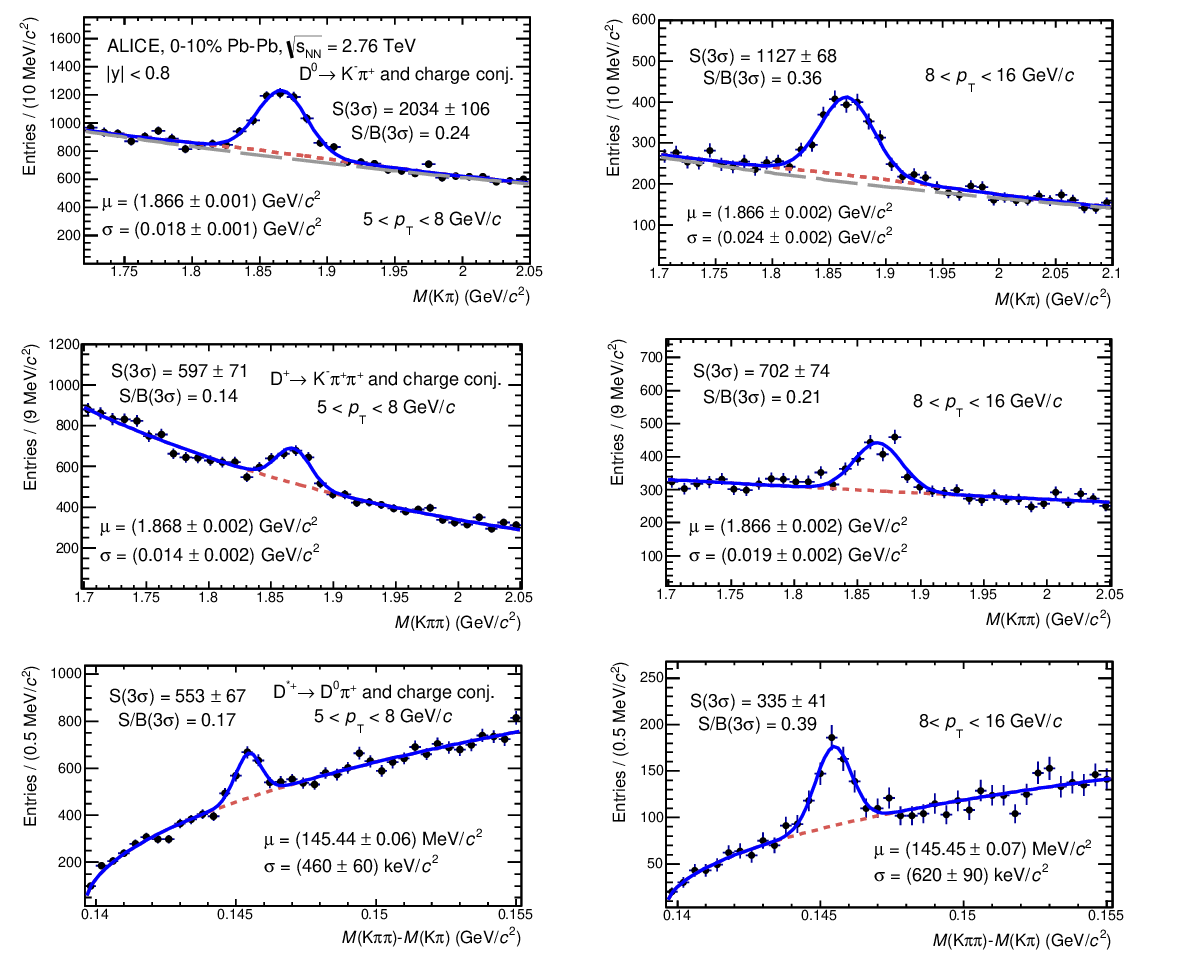}
  \caption{Distributions of the ${\rm K}\pi$ invariant mass for $\Dzero$ candidates (upper panels) and ${\rm K}\pi\pi$ invariant mass for $\Dplus$ candidates (central panels) and of the
invariant mass difference $M({\rm K}\pi\pi)-M({\rm K}\pi)$ for $\Dstar$ candidates (lower panels) and the corresponding charge conjugates in two $\pt$ intervals (left and right panels) for $16.4\times10^6$ Pb--Pb collisions in the 0--10\% centrality class. The curves show the fit functions described in the text. The red short-dashed line represents the background fit function.
For the $\Dzero$ meson, the gray dashed line represents the background without the inclusion of the template for the contribution of reflections, i.e.\,signal candidates with swapped ${\rm (K,\pi)}$ mass hypothesis. The template is defined as the sum of two Gaussians with parameters fixed to the values obtained in simulation.}
  \label{fig:MassPlots}
\end{figure}

In the most central centrality class (0--10\%), the statistical significance of the invariant mass signal peaks varies from 8 to 18 depending on the D meson species and $\pt$, while the signal-over-background ratio 
ranges from 0.1 to 0.4. In the most peripheral centrality class (50--80\%), the statistical significance varies from 4 to 11, while the signal-over-background ranges from 0.4 to 1.5.  In Fig.~\ref{fig:MassPlots} the invariant mass distributions of the three meson species are shown in the 0--10\% centrality class and in the transverse momentum intervals $5<\pt<8~\gev/c$ and $8<\pt<16~\gev/c$.

\begin{table}[!t]
\begin{center}
\begin{tabular}{lcccccc}
&\multicolumn{3}{c}{$5<\pt<8~\gev/c$}&\multicolumn{3}{c}{$8<\pt<16~\gev/c$} \\
\hline
& ~~~$\Dzero$~~~ & ~~~$\Dplus$~~~  & ~~~$\Dstar$~~~  &~~~$\Dzero$~~~ & ~~~$\Dplus$~~~ & ~~~$\Dstar$~~~ \\
\hline
Pb--Pb yields: & & &  & & &\\
~~~~Yield Extraction & 6 & 8 & 6  & 7 & 8 & 7  \\
~~~~Tracking efficiency & 10  & 15 & 15& 10 & 15 & 15 \\
~~~~PID identification & 5 & 5 & 5& 5 & 5 & 5 \\
~~~~Cut efficiency & 5 &  10 & 5& 5 & 10 & 5 \\
~~~~D $\pt$ distribution in sim. & 2 & 2 & 2& 2 & 2 & 2 \\
~~~~Feed-down subtraction & $^{+12}_{-13}$ & $^{+10}_{-10}$  & $^{+6}_{-8}$ & $^{+12}_{-12}$ & $^{+10}_{-10}$ & $^{+\phantom{1}7}_{-10}$ \\
$\av{\TAA}$~\cite{centrality} & \multicolumn{3}{c}{ 4}& \multicolumn{3}{c}{ 4} \\
pp reference &  16 &  20 &  17 & 16 & 19 & 17 \\
Reference scaling in $\sqrt{s}$ & \multicolumn{3}{c}{$^{+\phantom{1}6}_{-12}$} & \multicolumn{3}{c}{$^{+5}_{-6}$} \\
Centrality limits & \multicolumn{6}{c}{$<$ 0.1} \\
\hline
\end{tabular}
\caption{Systematic uncertainties (\%) on $\RAA$ of prompt D mesons with $5<\pt<8~\gev/c$ and $8<\pt<16~\gev/c$ in the 0--10\% centrality class.}
\label{tab:systematics}
\end{center}
\end{table}
The correction for acceptance and efficiency was determined using Monte Carlo simulations. Pb--Pb events were simulated using the HIJING 
generator~\cite{hijing} and D meson signals were added with the PYTHIA\,6 generator~\cite{pythia}. 
The $\pt$ distribution of the D mesons was weighted in order to match the shape measured for \mbox{$\rm D^0$ mesons} in central Pb--Pb collisions~\cite{DRAA2010}.
A detailed description of the detector response, based on 
the GEANT3 transport package~\cite{geant},  was included. The contribution of feed-down from ${\rm B\to D}+X$ to the inclusive D meson raw yield depends on $\pt$ and on the geometrical selection criteria, because the secondary vertices of \mbox{D mesons} from B-hadron decays are typically more displaced from the primary vertex than those of prompt D mesons. 
This contribution was subtracted using the beauty-hadron production cross section 
in pp collisions from FONLL calculations~\cite{FONLL}, convoluted with the decay kinematics as implemented in the EvtGen decay package~\cite{evtgen}
and multiplied by the efficiency for feed-down D mesons from the simulation, the average nuclear overlap function $\av{\TAA}$ in each centrality class, and an 
assumed value for $\RAA$ of feed-down D-mesons~\cite{DRAA2010}. 
On the basis of the comparison shown in this paper,  
 this assumption was taken as 
$\RAA^{\textnormal{feed-down D}} = 2\, \RAA^{\textnormal{prompt D}}$ and a systematic uncertainty was estimated 
by varying it in the interval $1<\RAA^{\textnormal{feed-down D}} / \RAA^{\textnormal{prompt D}} < 3$.
The feed-down contribution is about 20--25\%, depending on the D meson species and on the $\pt$ interval. 

 The $\pt$-differential cross section of prompt D mesons with $|y|<0.5$ in pp collisions at $\sqrt s=2.76~\tev$,
  used as reference for $\RAA$, 
  was obtained by scaling the measurement at $\sqrt{s}=7$~ TeV~\cite{Dpp7TeV}. The $\pt$-dependent scaling factor and its uncertainty were determined with FONLL 
  calculations~\cite{scaling}.  
  The result of the scaling was validated by comparison with the measurement obtained from a smaller sample of pp collisions 
  at $\sqrt{s}=2.76$ TeV~\cite{Dpp2.76TeV}. This measurement covers a reduced $\pt$ interval 1--12~$\gev/c$ with a statistical uncertainty of 20--25~$\%$ and was, therefore, not used as a pp reference in the present analysis.
 The yields in Pb--Pb collisions were normalized to the same rapidity interval as the reference ($|y|<0.5$) by dividing them by $\Delta y=1.6$.
 
The systematic uncertainties were estimated as a function of $\pt$ and centrality using the procedure described in~\cite{DRAA2010,ALICEDmesonv2article}
and briefly outlined in the following.
The sources of systematic uncertainty on the nuclear modification factor are listed in Table~\ref{tab:systematics}, along with their values for the two $\pt$ intervals in the most central collisions (0--10\%). 
The uncertainties are approximately independent of centrality.  

The systematic uncertainty on the yield extraction was estimated by varying the fit conditions (fit interval and functional form used to describe the background)
or by considering, as an alternative method, the bin counting of the invariant mass distribution obtained after subtracting the background estimated from a fit in the side-bands of the signal peak.
The uncertainty amounts to 
about  6--8\%. This includes in the case of the $\rm D^0$ a contribution of about 5\% obtained by varying the ratio of the integral of the reflections to the integral of the signal by $\pm~50\%$.

The systematic uncertainty on the tracking efficiency correction was evaluated by varying the track selection criteria and amounts to $5\%$ per track, thus 10\% for the $\rm D^0$ (two-track final state)
and 15\% for the $\rm D^+$ and $\rm D^{*+}$ mesons (three-track final states).
The correction for the particle identification (PID) efficiency introduces a systematic uncertainty of 5\%, which was estimated by repeating the analysis without this selection and comparing the
corrected yields.
A systematic uncertainty of  5--10$\%$ associated with the selection efficiency correction was estimated by varying the D meson selection cuts.
The D meson $\pt$ distribution used in the simulation to calculate the acceptance and efficiency was varied between the measured distribution and the prediction of a theoretical calculation including parton energy loss \cite{BAMPS, BAMPS1,Uphoff:2012gb}. The resulting variation of 2$\%$ of the efficiencies was assigned as a systematic uncertainty. 

The systematic uncertainty on the correction for feed-down from B-hadron decays was estimated, as described in~\cite{ALICEDmesonv2article}, by varying the parameters of the FONLL calculation and the hypothesis on the $\RAA$ of the feed-down D mesons in the range $1<\RAA^{\textnormal{feed-down D}} / \RAA^{\textnormal{prompt D}} < 3$. 
This variation yields the main contribution to the uncertainty,  which amounts to 6--13$\%$, depending on the D meson species and $\pt$ interval. 

The contribution to the systematic uncertainty due to the 1.1$\%$ relative uncertainty on the fraction of hadronic cross section used in the Glauber fit to determine the centrality classes was obtained as in \cite{DRAA2010} and estimated to be $<$ 0.1$\%$ in the central centrality class (0--10$\%$) and 3$\%$ in the most peripheral centrality class (50--80$\%$).

The systematic uncertainties on the denominator of the nuclear modification factor include the uncertainty on $\av{\TAA}$, which ranges from 4\% in the 0--10\% centrality class
to 7.5\% in the 50--80\% centrality class~\cite{centrality}, and the uncertainty on the pp reference. The latter has a contribution of about 16--20\% from the pp measurement at $\sqrt s=7~\tev$
and a contribution of $^{+12}_{-\phantom{1}6}\%$ from the energy scaling down to $\sqrt s=2.76~\tev$.

\section{Results and discussion}
\label{sec:DRAA}
\begin{figure*}
  \centering
  \includegraphics[width=0.48\textwidth]{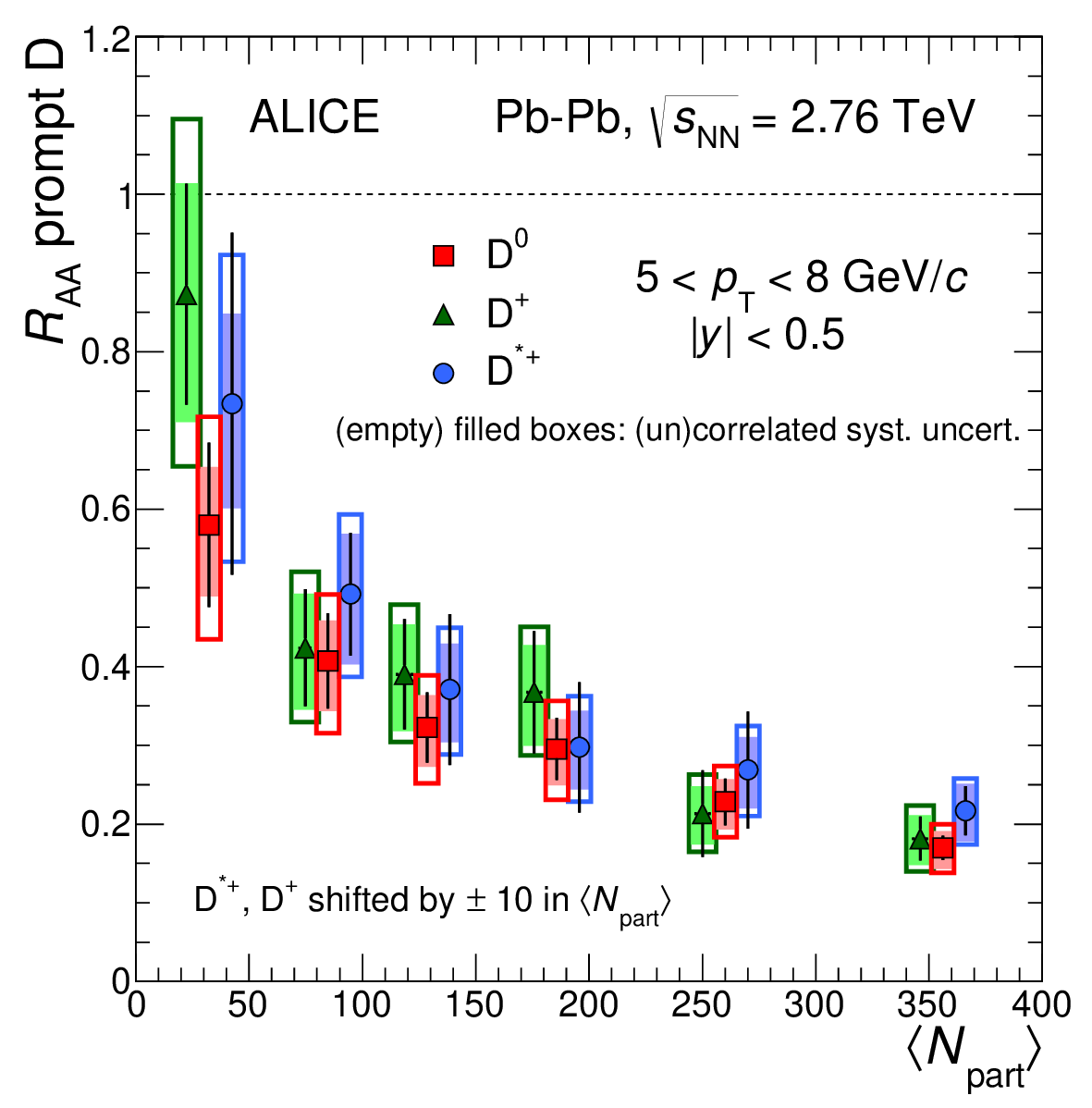}
  \includegraphics[width=0.48\textwidth]{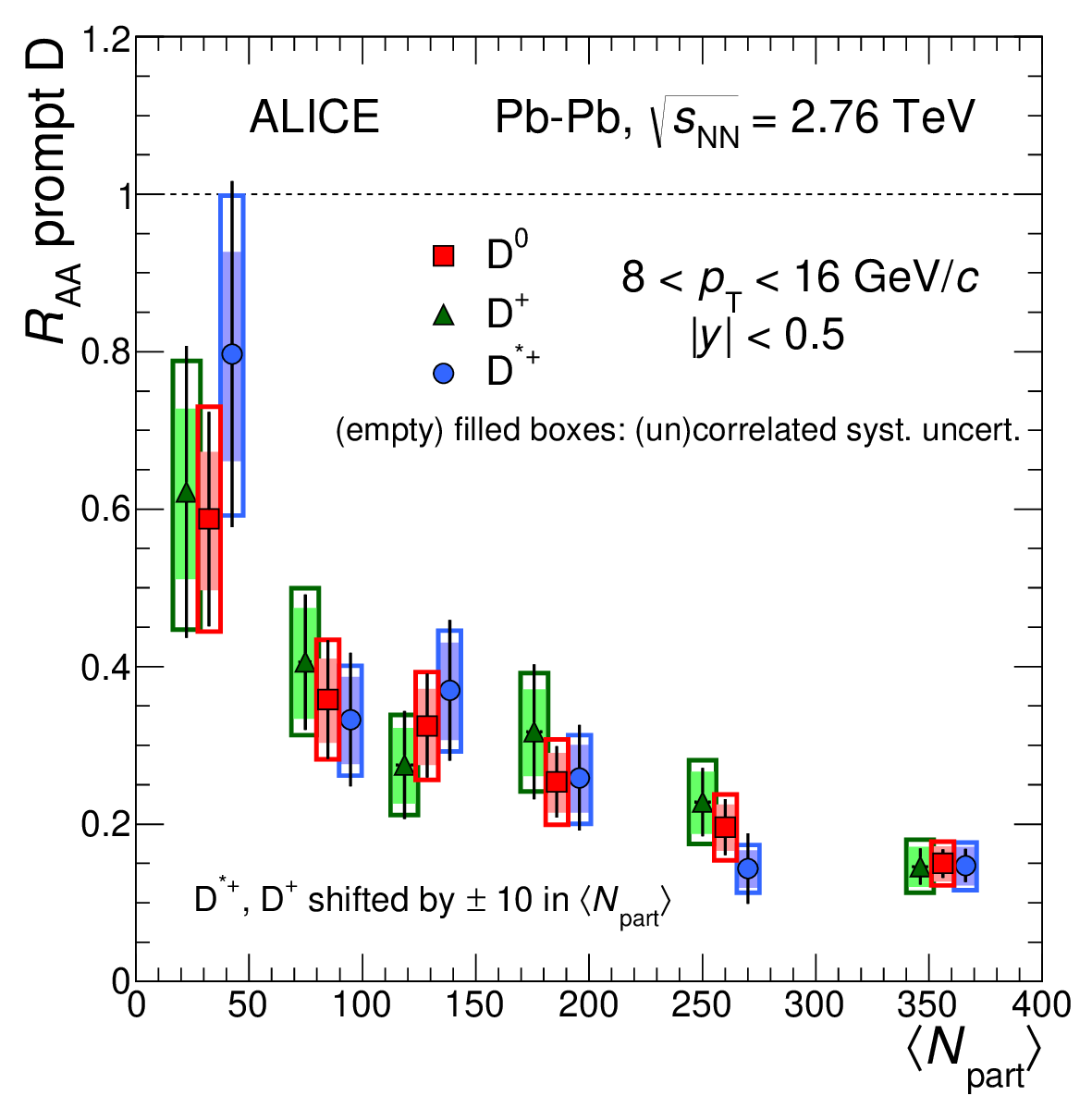}
  \caption{$\RAA$ as a function of centrality 
    ($\langle N_{\rm part}\rangle$, see text)  
of $\Dzero$, $\Dplus$ and $\Dstar$ in $5<\pt<8~\gev/c$ (left) and $8<\pt<16~\gev/c$ 
    (right). The bars represent the statistical uncertainty while the filled (empty) 
    boxes represent the systematic uncertainties that are correlated (uncorrelated) among centrality intervals. The symbols for $\Dstar$ and $\Dplus$ are shifted by $\langle N_{\rm part}\rangle$ = 10 for better visibility.}
  \label{fig:RAA2ptbins}
\end{figure*}

Figure~\ref{fig:RAA2ptbins} shows the $\RAA$ as a function of centrality 
for $\Dzero$, $\Dplus$ and $\Dstar$ in the intervals $5<\pt<8~\gev/c$ (left) 
and $8<\pt<16~\gev/c$ (right). 
Centrality is quantified in terms of 
the average number of nucleons participating in the collision in each multiplicity class, $\langle {N_{\rm part}}\rangle$, evaluated with a Monte Carlo Glauber calculation~\cite{centrality}.
The bars represent the statistical uncertainties.
The filled and empty boxes represent the quadratic sum of the systematic uncertainties
that are, respectively, correlated between centrality intervals (pp reference, \mbox{B-hadron cross section} used for feed-down 
correction, particle identification, track reconstruction efficiency, $\langle\TAA\rangle$) and uncorrelated 
(yield extraction, selection efficiency corrections, value of feed-down \mbox{D meson $\RAA$}). 
The latter category also includes the systematic uncertainties that are partially correlated between adjacent centrality classes.
The measurements for the three D meson species share part of the systematic uncertainties and are consistent within statistical uncertainties.
 The suppression increases with centrality and reaches a factor of 5--6 in the most central collisions for both $\pt$ intervals.  

A weighted average of the $\RAA$ of the three D meson species was computed using the 
 inverse of the relative statistical uncertainties as weights.
 The systematic uncertainties of the weighted average were calculated 
 considering the contributions from the tracking efficiency, 
the feed-down correction, and the reference energy scaling factor from 7 to 
2.76~TeV as fully correlated among the three D meson species. 

\begin{figure*}
  \centering
  \includegraphics[width=0.48\textwidth]{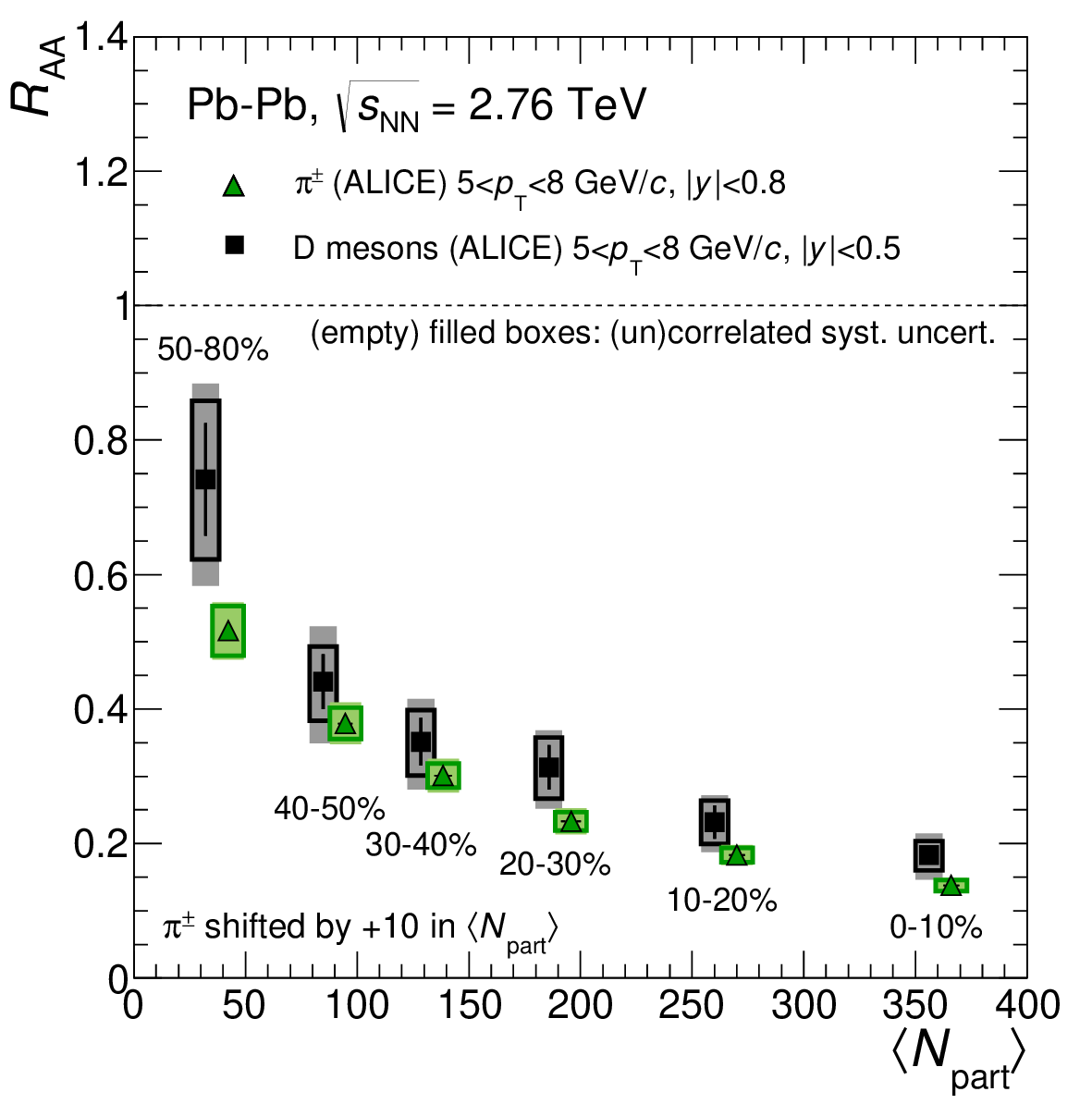}
  \includegraphics[width=0.48\textwidth]{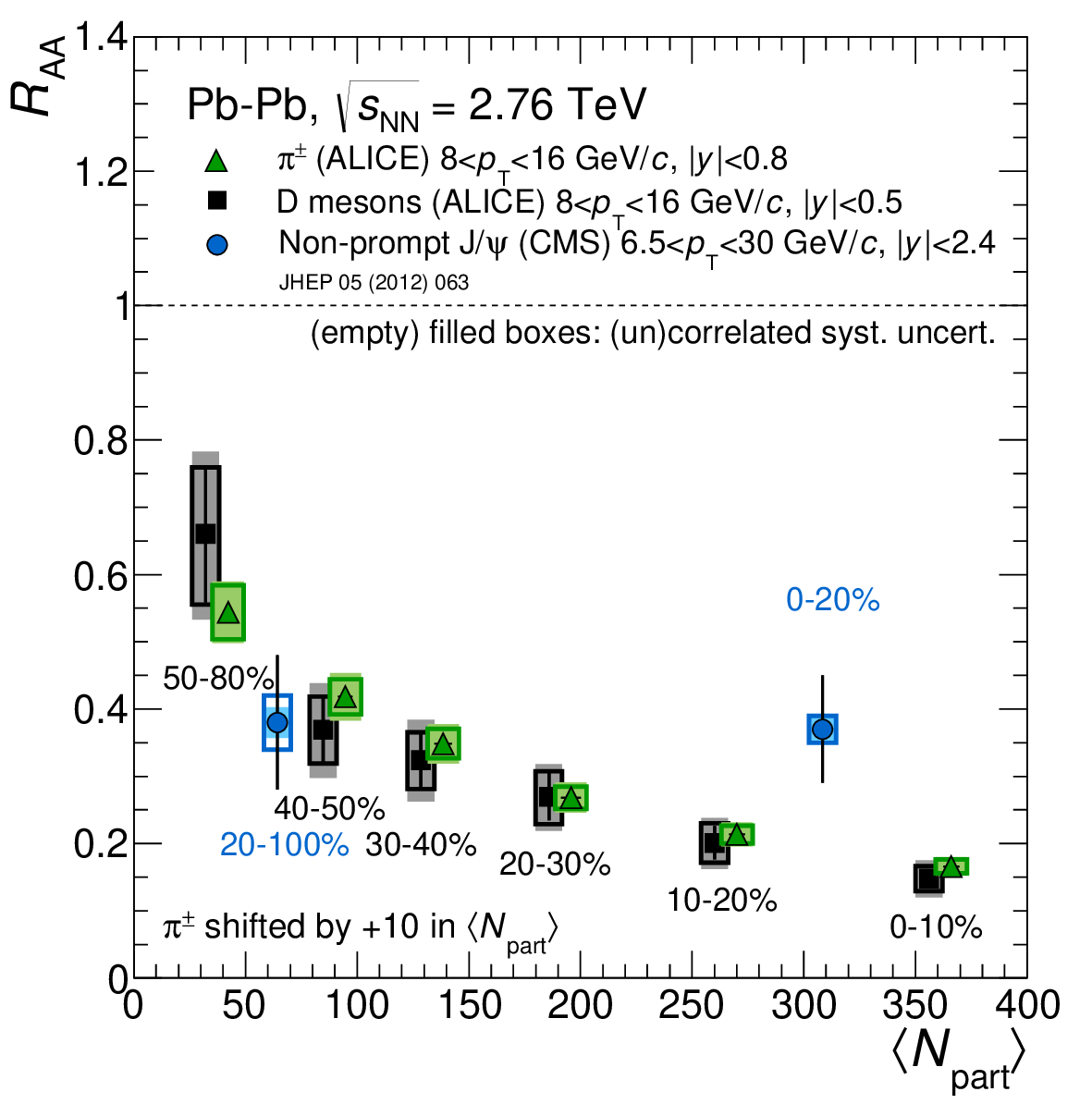}
  \caption{Comparison of the D meson $\RAA$ (average of $\Dzero$, $\Dplus$ and $\Dstar$) and  
    of the charged pion $\RAA$~\cite{ALICEhighptPIDRAA} in \mbox{$5<\pt<8~\gev/c$} (left) and in $8<\pt<16~\gev/c$ (right). The right panel also includes 
    the $\RAA$ of non-prompt $\Jpsi$ mesons in $6.5<\pt<30~\gev/c$ measured by the CMS Collaboration~\cite{CMSJpsi2010}.
    The vertical bars represent the statistical uncertainties. The D meson systematic uncertainties are displayed as in the previous figures.
    The total systematic uncertainties of charged pions are shown by boxes.
    The centrality-dependent systematic uncertainties are shown by boxes on the individual data points.}
  \label{fig:RAAcomparisonCMSjpsi}
\end{figure*}

Figure~\ref{fig:RAAcomparisonCMSjpsi} shows the average of the $\Dzero$, $\Dplus$
and $\Dstar$ nuclear modification factors as a function of centrality, for the
intervals $5<\pt<8~\gev/c$ (left) and $8<\pt<16~\gev/c$ (right), compared with
the $\RAA$ of charged pions with $|y|<0.8$ for the same $\pt$ intervals\footnote{The charged pion results were obtained with the analysis method described in~\cite{ALICEhighptPIDRAA}.}, and of
non-prompt $\Jpsi$ mesons
measured by the CMS Collaboration for $6.5<\pt<30~\gev/c$ 
in $|y|<2.4$~\cite{CMSJpsi2010}. Care has to be taken when comparing with the non-central CMS data point as it is plotted at the $\Npart$ mean value of the broad 20--100$\%$ centrality interval.

The $\pt$ interval 8--16~GeV/$c$ for D mesons was chosen in order to obtain a significant overlap
with the $\pt$ distribution of B mesons decaying to $\Jpsi$ particles with $6.5<\pt<30~\gev/c$.
Using a simulation based on the FONLL calculation~\cite{FONLL} and the 
EvtGen particle decay package~\cite{evtgen}, it was estimated that about
70\% of these parent B mesons have $8<\pt<16~\gev/c$, with a median of the $\pt$ distribution
of about 11.3~$\gev/c$. A median value of $(9.5~\pm ~0.5)~\gev/c$ was estimated for D mesons 
with $8<\pt<16~\gev/c$ in the 0--10\% centrality class. The estimate was based on the $\pt$ distribution of 
$\rm D^0$ mesons in $\pt$ intervals with a width of $1~\gev/c$. 
The effect of the different width of the rapidity interval for D and non-prompt $\Jpsi$ mesons ($|y|<0.5$ and $|y|<2.4$, respectively)
is expected to be mild because the intervals are partially overlapping and a preliminary measurement by the CMS Collaboration does not indicate a significant $y$ dependence of the $\RAA$ of non-prompt $\Jpsi$ mesons in $|y|<2.4$~\cite{CMSJpsi2011}.

The nuclear modification factors of charged pions and D mesons are compatible within 
uncertainties in all centrality classes and in the two $\pt$ intervals. The value of the D meson $\RAA$ 
in the centrality classes 0--10\% and 10--20\% for $8<\pt<16~\gev/c$ is lower than that of
non-prompt $\Jpsi$ mesons in the centrality class 0--20\%.  However, the difference between the $\RAA$ values
is not larger than $3\,\sigma$, considering the statistical and systematic uncertainties.
A preliminary higher-statistics measurement by the CMS Collaboration 
of non-prompt J/$\psi$ production in the same $\pt$ interval (6.5--30~$\gev/c$) 
and in a narrower rapidity interval $(|y|<1.2)$ is also available~\cite{CMSJpsi2011}.
Considering this measurement,
the average difference of the $\RAA$ values of D mesons and non-prompt $\Jpsi$
in the 0--10\% and 10--20\% centrality classes is larger than zero 
with a significance of 3.5 $\sigma$, obtained including the systematic uncertainties, 
and taking into account their correlation between 
the two centrality classes. 

\begin{figure}
  \centering
  \includegraphics[width=0.49\textwidth]{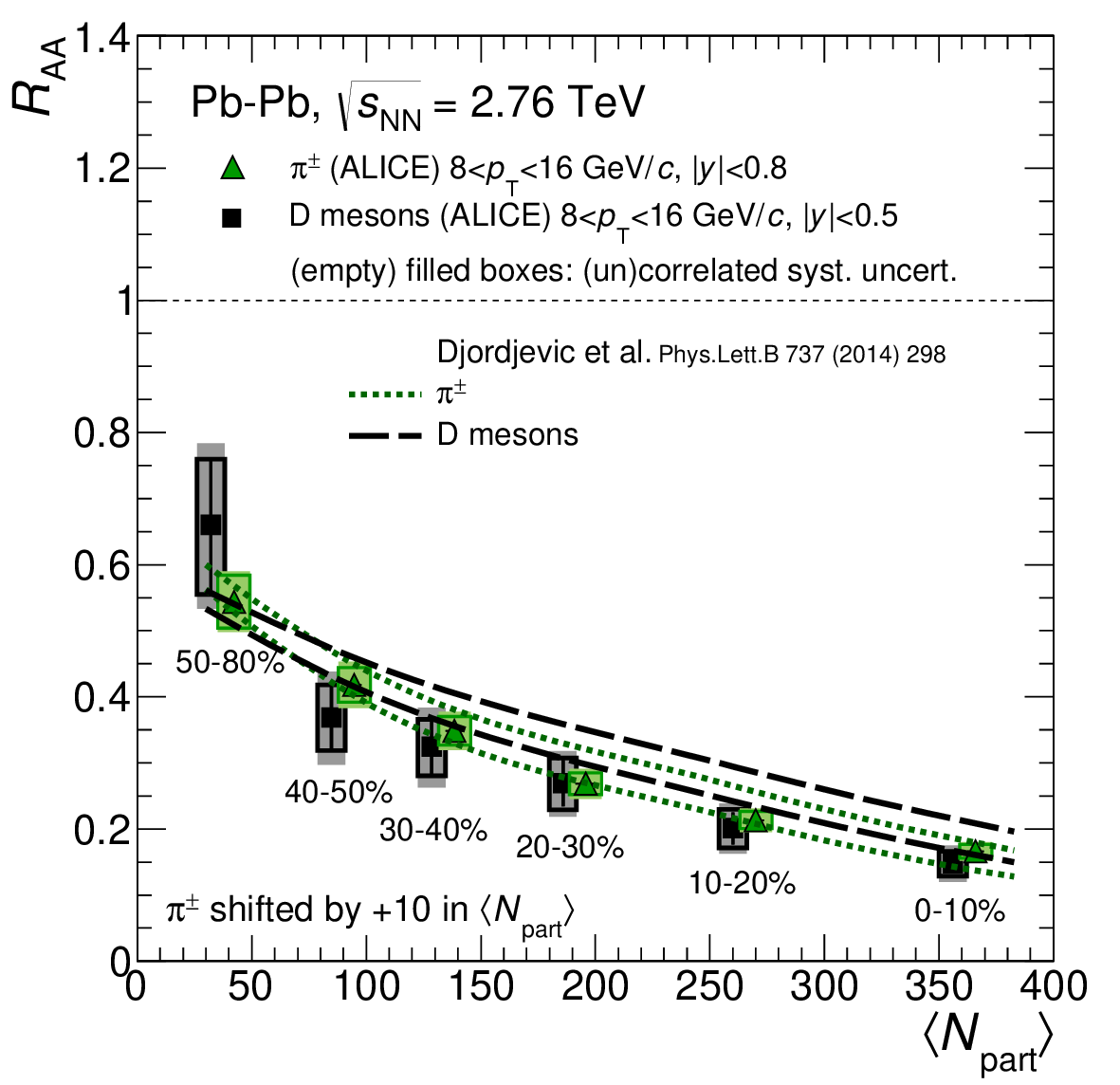}
  \includegraphics[width=0.49\textwidth]{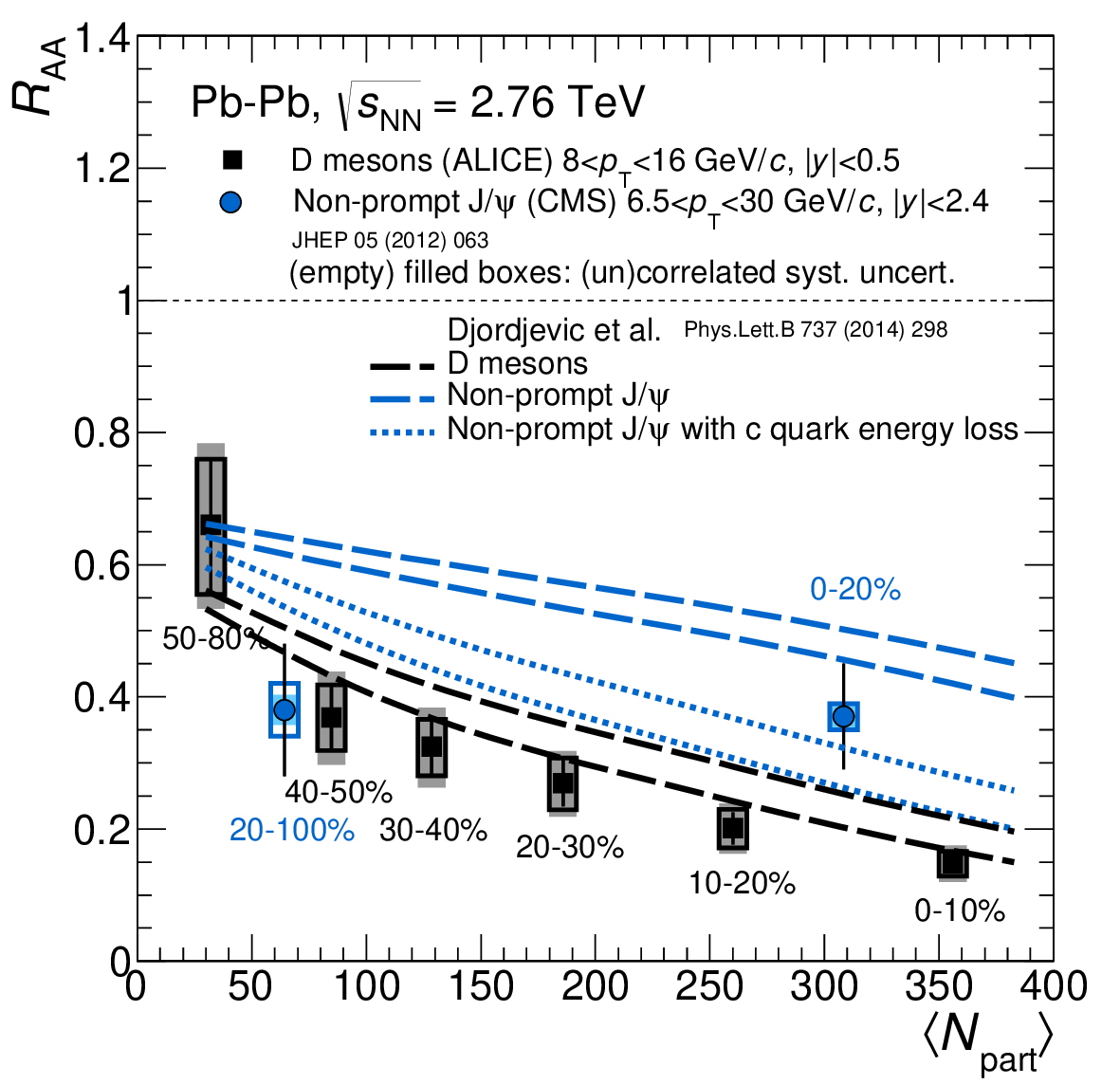}
  \caption{Comparison of the $\RAA$ measurements with the calculations by Djordjevic {\it et al.}~\cite{Djordjevic:2014tka} including radiative and collisional 
  energy loss. Lines of the same style enclose a band representing the theoretical uncertainty.
Left: D mesons and charged pions in $8<\pt<16~\gev/c$. Right: D mesons in $8<\pt<16~\gev/c$ and non-prompt J/$\psi$ mesons in $6.5<\pt<30~\gev/c$~\cite{CMSJpsi2010}. 
For the latter, the model results for the case in which the b quark interactions are calculated 
using the c quark mass are shown as well~\cite{SaporeGravis}. }
  \label{fig:RAAcompDjordjevic}
\end{figure}

\begin{figure}
  \centering
  \includegraphics[width=0.49\textwidth]{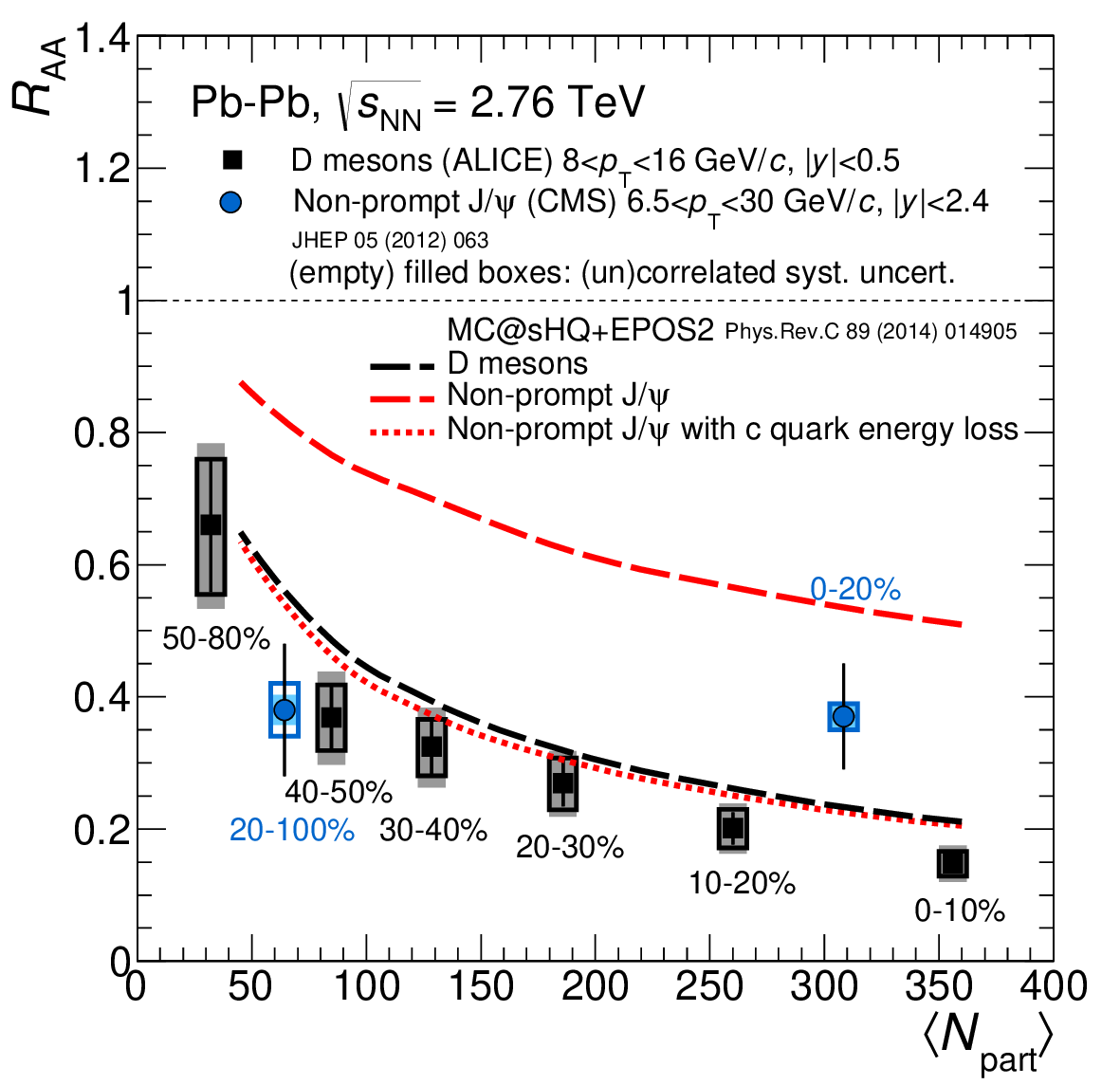}
  \includegraphics[width=0.49\textwidth]{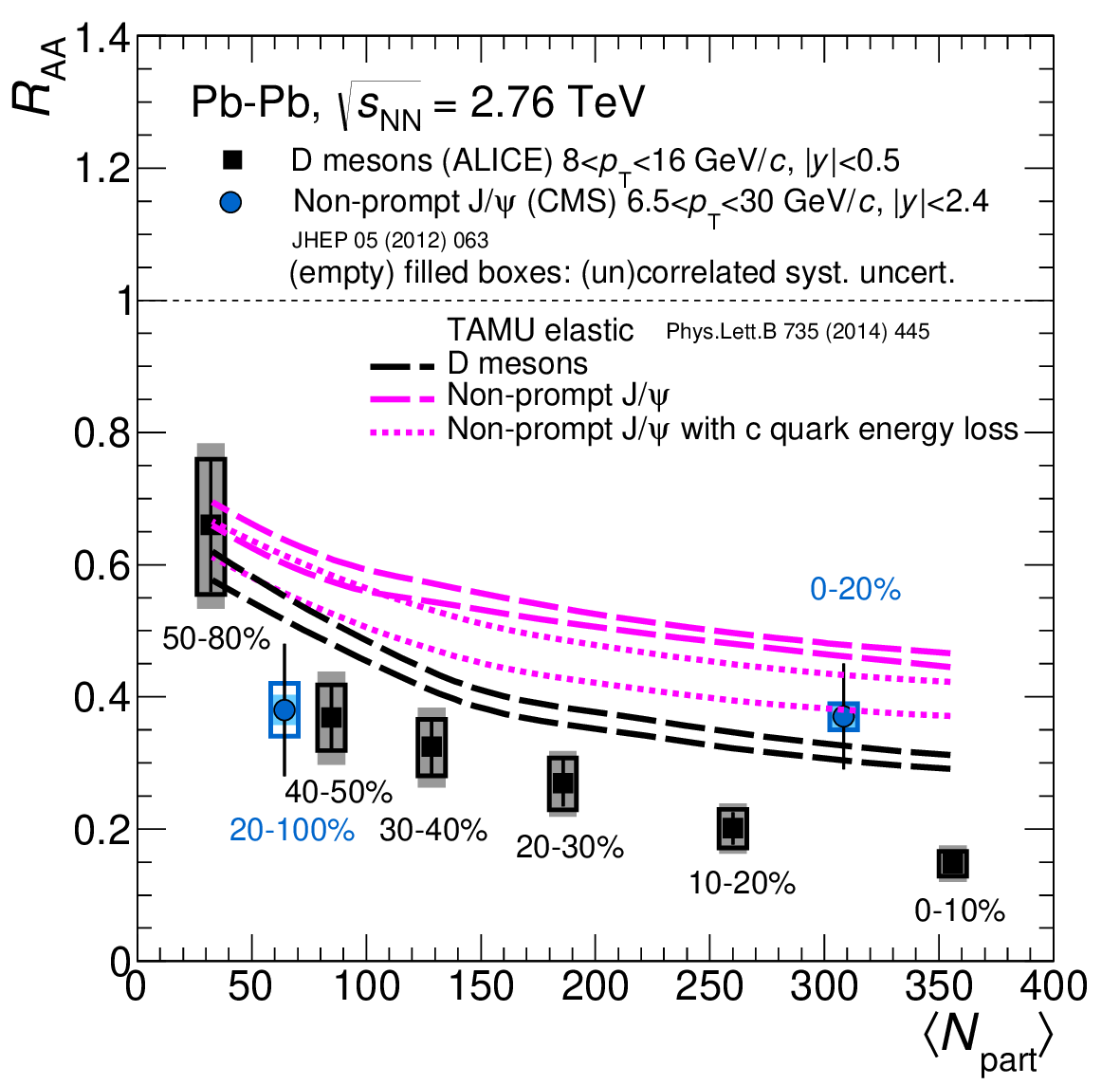}
  \caption{Comparison of the $\RAA$ measurements for D mesons ($8<\pt<16~\gev/c$) and non-prompt J/$\psi$ mesons ($6.5<\pt<30~\gev/c$)~\cite{CMSJpsi2010} with the MC@sHQ+EPOS2 model~\cite{gossiauxEPOS} including radiative and collisional 
  interactions (left) and with the TAMU elastic model~\cite{He:2012xz} including collisional interactions via in-medium resonance formation.
For both models, results for the case in which the b quark interactions are calculated using the c quark mass are shown as well~\cite{SaporeGravis}. 
In the right-hand panel, the band between lines with the same style represents the theoretical uncertainty.}
  \label{fig:RAAcompGossiauxRapp}
\end{figure}

The nuclear modification factors of D mesons (average of $\Dzero$, $\Dplus$ and $\Dstar$) and charged pions
in the interval $8<\pt<16~\gev/c$ and that of non-prompt J/$\psi$ mesons in $6.5<\pt<30~\gev/c$ were compared with
theoretical calculations. Figure~\ref{fig:RAAcompDjordjevic} shows the comparison with the calculation by Djordjevic {\it et al.}~\cite{Djordjevic:2014tka}.
This model implements energy loss for gluons, light and heavy quarks, including both radiative (DGLV formalism~\cite{dg}) and collisional processes and 
considers dynamical scattering centres in the medium. The heavy-quark production $\pt$-differential cross sections
are obtained from FONLL calculations~\cite{FONLL} and hadronization assumes fragmentation outside the medium.
In the left-hand panel, the calculation closely describes the similarity of the D meson and charged pion $\RAA$ over the entire centrality range.
As mentioned in the introduction, in this calculation the colour-charge dependence of energy loss introduces a sizeable difference in the 
suppression of the gluon and c quark production. However, the softer fragmentation and $\pt$ spectrum of gluons with respect to those of c quarks,
together with the increase of the parton-level $\RAA$ with increasing $\pt$, lead to a compensation effect that results in 
a very similar $\RAA$ for D mesons and pions~\cite{Djordjevic:2013pba}.
As shown in the right-hand panel of the figure, this calculation results in a larger suppression of D mesons with respect to non-prompt J/$\psi$, 
in qualitative agreement with the data for the most central collisions. 
In order to study the origin of this large difference in the calculation, the result for a test case with the energy loss of b quarks calculated 
using the c quark mass was considered~\cite{SaporeGravis}. 
In this case, the $\RAA$ of non-prompt J/$\psi$
was found to be quite close to that of D mesons. This indicates that, in the calculation, the large difference in the $\RAA$ of 
D mesons and non-prompt J/$\psi$ derives predominantly from the quark mass dependence of the parton energy loss.

In Fig.~\ref{fig:RAAcompGossiauxRapp} the D meson and non-prompt J/$\psi$ data are compared with two theoretical models that implement heavy-quark 
interactions in an expanding hydrodynamical medium. The MC@sHQ+EPOS2 model~\cite{gossiauxEPOS}, shown in the left-hand panel, includes radiative and collisional energy loss.
The hydrodynamical evolution of the medium is simulated using the EPOS2 model~\cite{EPOS,EPOS1}. Heavy-quark transport in the medium is based on the 
Boltzmann equation, with collisional processes and radiative corrections. 
The TAMU elastic model~\cite{He:2012xz}, shown in the right-hand panel, includes collisional (elastic) processes only. 
In this model, the heavy-quark transport coefficient is calculated within a non-perturbative $T$-matrix approach,
where the interactions proceed via resonance formation that transfers momentum from the heavy quarks to the medium constituents. 
The model includes hydrodynamic medium evolution, constrained by light-flavour hadron production data.
Elastic diffusion of heavy-flavour hadrons in the hadronic phase is included as well. 
In both models, similarly to that of Djordjevic {\it et al.}, the heavy-quark production cross sections
are obtained from the FONLL calculation~\cite{FONLL}. 
Both models implement a contribution of quark recombination in the hadronization of heavy quarks, in addition to 
fragmentation outside the medium.
The dotted lines correspond to the test case in which the b quark mass is decreased to the c quark mass value in the calculation of the in-medium interactions~\cite{SaporeGravis}.

The MC@sHQ+EPOS2 model qualitatively describes the two measurements in these $\pt$ intervals.
In this model a large difference in the suppression of D mesons and non-prompt J/$\psi$ is caused by the mass dependence of energy loss as in Djordjevic {\it et al.} model.
The TAMU elastic model tends to overestimate $\RAA$ for both the non-prompt J/$\psi$ and the D mesons, in particular in central collisions. 
At variance with the other two models, in this case the quark mass effect accounts for only about half of the difference in the suppression of D and non-prompt J/$\psi$ mesons.
This model does not include radiative energy loss, which is expected to have a strong mass dependence.

The nuclear modification factors of D mesons and non-prompt J/$\psi$ are also described by a model calculation 
by the Duke group~\cite{Cao:2015hia}, that includes radiative and collisional energy loss within an hydro- dynamical medium
and performs the hadronization of heavy quarks using recombination and   fragmentation.

\section{Summary}
\label{sec:conclusions}
The centrality dependence of the nuclear modification factor of prompt D mesons
in $\PbPb$ collisions at $\sqrts = 2.76~\tev$ was presented
in the intervals $5<\pt<8~\gev/c$ and $8<\pt<16~\gev/c$.
A suppression is observed already in the centrality class 50--80\% and it increases
towards more central collisions, reaching a maximum of a factor about 5--6 in the most central collisions.

The centrality dependence and the magnitude of the suppression are similar to those of charged pions 
in the same $\pt$ intervals.
The comparison of the D meson $\RAA$ with the non-prompt J/$\psi$ meson $\RAA$ hints at a difference 
in the suppression of particles originating from c and b quarks in the most central collisions. 

These results are described by theoretical calculations 
in which in-medium parton energy loss increases with increasing colour charge factor and 
decreases with increasing quark mass. Calculations that include radiative energy loss,
in addition to collisional energy loss, provide a better quantitative description of 
the data.


%
%
\newenvironment{acknowledgement}{\relax}{\relax}
\section*{Acknowledgements}

The ALICE Collaboration would like to thank all its engineers and technicians for their invaluable contributions to the construction of the experiment and the CERN accelerator teams for the outstanding performance of the LHC complex.
The ALICE Collaboration gratefully acknowledges the resources and support provided by all Grid centres and the Worldwide LHC Computing Grid (WLCG) collaboration.
The ALICE Collaboration acknowledges the following funding agencies for their support in building and
running the ALICE detector:
State Committee of Science,  World Federation of Scientists (WFS)
and Swiss Fonds Kidagan, Armenia,
Conselho Nacional de Desenvolvimento Cient\'{\i}fico e Tecnol\'{o}gico (CNPq), Financiadora de Estudos e Projetos (FINEP),
Funda\c{c}\~{a}o de Amparo \`{a} Pesquisa do Estado de S\~{a}o Paulo (FAPESP);
National Natural Science Foundation of China (NSFC), the Chinese Ministry of Education (CMOE)
and the Ministry of Science and Technology of China (MSTC);
Ministry of Education and Youth of the Czech Republic;
Danish Natural Science Research Council, the Carlsberg Foundation and the Danish National Research Foundation;
The European Research Council under the European Community's Seventh Framework Programme;
Helsinki Institute of Physics and the Academy of Finland;
French CNRS-IN2P3, the `Region Pays de Loire', `Region Alsace', `Region Auvergne' and CEA, France;
German Bundesministerium fur Bildung, Wissenschaft, Forschung und Technologie (BMBF) and the Helmholtz Association;
General Secretariat for Research and Technology, Ministry of
Development, Greece;
Hungarian Orszagos Tudomanyos Kutatasi Alappgrammok (OTKA) and National Office for Research and Technology (NKTH);
Department of Atomic Energy and Department of Science and Technology of the Government of India;
Istituto Nazionale di Fisica Nucleare (INFN) and Centro Fermi -
Museo Storico della Fisica e Centro Studi e Ricerche "Enrico
Fermi", Italy;
MEXT Grant-in-Aid for Specially Promoted Research, Ja\-pan;
Joint Institute for Nuclear Research, Dubna;
National Research Foundation of Korea (NRF);
Consejo Nacional de Cienca y Tecnologia (CONACYT), Direccion General de Asuntos del Personal Academico(DGAPA), M\'{e}xico, Amerique Latine Formation academique - European Commission~(ALFA-EC) and the EPLANET Program~(European Particle Physics Latin American Network);
Stichting voor Fundamenteel Onderzoek der Materie (FOM) and the Nederlandse Organisatie voor Wetenschappelijk Onderzoek (NWO), Netherlands;
Research Council of Norway (NFR);
National Science Centre, Poland;
Ministry of National Education/Institute for Atomic Physics and National Council of Scientific Research in Higher Education~(CNCSI-UEFISCDI), Romania;
Ministry of Education and Science of Russian Federation, Russian
Academy of Sciences, Russian Federal Agency of Atomic Energy,
Russian Federal Agency for Science and Innovations and The Russian
Foundation for Basic Research;
Ministry of Education of Slovakia;
Department of Science and Technology, South Africa;
Centro de Investigaciones Energeticas, Medioambientales y Tecnologicas (CIEMAT), E-Infrastructure shared between Europe and Latin America (EELA), Ministerio de Econom\'{i}a y Competitividad (MINECO) of Spain, Xunta de Galicia (Conseller\'{\i}a de Educaci\'{o}n),
Centro de Aplicaciones Tecnológicas y Desarrollo Nuclear (CEA\-DEN), Cubaenerg\'{\i}a, Cuba, and IAEA (International Atomic Energy Agency);
Swedish Research Council (VR) and Knut $\&$ Alice Wallenberg
Foundation (KAW);
Ukraine Ministry of Education and Science;
United Kingdom Science and Technology Facilities Council (STFC);
The United States Department of Energy, the United States National
Science Foundation, the State of Texas, and the State of Ohio;
Ministry of Science, Education and Sports of Croatia and  Unity through Knowledge Fund, Croatia.
Council of Scientific and Industrial Research (CSIR), New Delhi, India

\bibliographystyle{utphys}	
\bibliography{biblio.bib}

\newpage
\appendix

\section{ALICE Collaboration}
\label{app:collab} 



\begingroup
\small
\begin{flushleft}
J.~Adam\Irefn{org40}\And
D.~Adamov\'{a}\Irefn{org83}\And
M.M.~Aggarwal\Irefn{org87}\And
G.~Aglieri Rinella\Irefn{org36}\And
M.~Agnello\Irefn{org111}\And
N.~Agrawal\Irefn{org48}\And
Z.~Ahammed\Irefn{org132}\And
S.U.~Ahn\Irefn{org68}\And
I.~Aimo\Irefn{org94}\textsuperscript{,}\Irefn{org111}\And
S.~Aiola\Irefn{org137}
% ...
\renewcommand\labelenumi{\textsuperscript{\theenumi}~}

\section*{Affiliation notes}
\renewcommand\theenumi{\roman{enumi}}
\begin{Authlist}
\item \Adef{0}Deceased
\item \Adef{idp3797616}{Also at: M.V. Lomonosov Moscow State University, D.V. Skobeltsyn Institute of Nuclear, Physics, Moscow, Russia}
\item \Adef{idp5933408}{Also at: University of Kansas, Lawrence, Kansas, United States}
\end{Authlist}

\section*{Collaboration Institutes}
\renewcommand\theenumi{\arabic{enumi}~}
\begin{Authlist}

\item \Idef{org1}A.I. Alikhanyan National Science Laboratory (Yerevan Physics Institute) Foundation, Yerevan, Armenia
\item \Idef{org2}Benem\'{e}rita Universidad Aut\'{o}noma de Puebla, Puebla, Mexico
\item \Idef{org3}Bogolyubov Institute for Theoretical Physics, Kiev, Ukraine
\item \Idef{org4}Bose Institute, Department of Physics and Centre for Astroparticle Physics and Space Science (CAPSS), Kolkata, India
\item \Idef{org5}Budker Institute for Nuclear Physics, Novosibirsk, Russia
\item \Idef{org6}California Polytechnic State University, San Luis Obispo, California, United States
\item \Idef{org7}Central China Normal University, Wuhan, China
\item \Idef{org8}Centre de Calcul de l'IN2P3, Villeurbanne, France
\item \Idef{org9}Centro de Aplicaciones Tecnol\'{o}gicas y Desarrollo Nuclear (CEADEN), Havana, Cuba
\item \Idef{org10}Centro de Investigaciones Energ\'{e}ticas Medioambientales y Tecnol\'{o}gicas (CIEMAT), Madrid, Spain
\item \Idef{org11}Centro de Investigaci\'{o}n y de Estudios Avanzados (CINVESTAV), Mexico City and M\'{e}rida, Mexico
\item \Idef{org12}Centro Fermi - Museo Storico della Fisica e Centro Studi e Ricerche ``Enrico Fermi'', Rome, Italy
\item \Idef{org13}Chicago State University, Chicago, Illinois, USA
\item \Idef{org14}China Institute of Atomic Energy, Beijing, China
\item \Idef{org15}Commissariat \`{a} l'Energie Atomique, IRFU, Saclay, France
\item \Idef{org16}COMSATS Institute of Information Technology (CIIT), Islamabad, Pakistan
\item \Idef{org17}Departamento de F\'{\i}sica de Part\'{\i}culas and IGFAE, Universidad de Santiago de Compostela, Santiago de Compostela, Spain
\item \Idef{org18}Department of Physics and Technology, University of Bergen, Bergen, Norway
\item \Idef{org19}Department of Physics, Aligarh Muslim University, Aligarh, India
\item \Idef{org20}Department of Physics, Ohio State University, Columbus, Ohio, United States
\item \Idef{org21}Department of Physics, Sejong University, Seoul, South Korea
\item \Idef{org22}Department of Physics, University of Oslo, Oslo, Norway
\item \Idef{org23}Dipartimento di Elettrotecnica ed Elettronica del Politecnico, Bari, Italy
\item \Idef{org24}Dipartimento di Fisica dell'Universit\`{a} 'La Sapienza' and Sezione INFN Rome, Italy
\item \Idef{org25}Dipartimento di Fisica dell'Universit\`{a} and Sezione INFN, Cagliari, Italy
\item \Idef{org26}Dipartimento di Fisica dell'Universit\`{a} and Sezione INFN, Trieste, Italy
\item \Idef{org27}Dipartimento di Fisica dell'Universit\`{a} and Sezione INFN, Turin, Italy
\item \Idef{org28}Dipartimento di Fisica e Astronomia dell'Universit\`{a} and Sezione INFN, Bologna, Italy
\item \Idef{org29}Dipartimento di Fisica e Astronomia dell'Universit\`{a} and Sezione INFN, Catania, Italy
\item \Idef{org30}Dipartimento di Fisica e Astronomia dell'Universit\`{a} and Sezione INFN, Padova, Italy
\item \Idef{org31}Dipartimento di Fisica `E.R.~Caianiello' dell'Universit\`{a} and Gruppo Collegato INFN, Salerno, Italy
\item \Idef{org32}Dipartimento di Scienze e Innovazione Tecnologica dell'Universit\`{a} del  Piemonte Orientale and Gruppo Collegato INFN, Alessandria, Italy
\item \Idef{org33}Dipartimento Interateneo di Fisica `M.~Merlin' and Sezione INFN, Bari, Italy
\item \Idef{org34}Division of Experimental High Energy Physics, University of Lund, Lund, Sweden
\item \Idef{org35}Eberhard Karls Universit\"{a}t T\"{u}bingen, T\"{u}bingen, Germany
\item \Idef{org36}European Organization for Nuclear Research (CERN), Geneva, Switzerland
\item \Idef{org37}Excellence Cluster Universe, Technische Universit\"{a}t M\"{u}nchen, Munich, Germany
\item \Idef{org38}Faculty of Engineering, Bergen University College, Bergen, Norway
\item \Idef{org39}Faculty of Mathematics, Physics and Informatics, Comenius University, Bratislava, Slovakia
\item \Idef{org40}Faculty of Nuclear Sciences and Physical Engineering, Czech Technical University in Prague, Prague, Czech Republic
\item \Idef{org41}Faculty of Science, P.J.~\v{S}af\'{a}rik University, Ko\v{s}ice, Slovakia
\item \Idef{org42}Faculty of Technology, Buskerud and Vestfold University College, Vestfold, Norway
\item \Idef{org43}Frankfurt Institute for Advanced Studies, Johann Wolfgang Goethe-Universit\"{a}t Frankfurt, Frankfurt, Germany
\item \Idef{org44}Gangneung-Wonju National University, Gangneung, South Korea
\item \Idef{org45}Gauhati University, Department of Physics, Guwahati, India
\item \Idef{org46}Helsinki Institute of Physics (HIP), Helsinki, Finland
\item \Idef{org47}Hiroshima University, Hiroshima, Japan
\item \Idef{org48}Indian Institute of Technology Bombay (IIT), Mumbai, India
\item \Idef{org49}Indian Institute of Technology Indore, Indore (IITI), India
\item \Idef{org50}Inha University, Incheon, South Korea
\item \Idef{org51}Institut de Physique Nucl\'eaire d'Orsay (IPNO), Universit\'e Paris-Sud, CNRS-IN2P3, Orsay, France
\item \Idef{org52}Institut f\"{u}r Informatik, Johann Wolfgang Goethe-Universit\"{a}t Frankfurt, Frankfurt, Germany
\item \Idef{org53}Institut f\"{u}r Kernphysik, Johann Wolfgang Goethe-Universit\"{a}t Frankfurt, Frankfurt, Germany
\item \Idef{org54}Institut f\"{u}r Kernphysik, Westf\"{a}lische Wilhelms-Universit\"{a}t M\"{u}nster, M\"{u}nster, Germany
\item \Idef{org55}Institut Pluridisciplinaire Hubert Curien (IPHC), Universit\'{e} de Strasbourg, CNRS-IN2P3, Strasbourg, France
\item \Idef{org56}Institute for Nuclear Research, Academy of Sciences, Moscow, Russia
\item \Idef{org57}Institute for Subatomic Physics of Utrecht University, Utrecht, Netherlands
\item \Idef{org58}Institute for Theoretical and Experimental Physics, Moscow, Russia
\item \Idef{org59}Institute of Experimental Physics, Slovak Academy of Sciences, Ko\v{s}ice, Slovakia
\item \Idef{org60}Institute of Physics, Academy of Sciences of the Czech Republic, Prague, Czech Republic
\item \Idef{org61}Institute of Physics, Bhubaneswar, India
\item \Idef{org62}Institute of Space Science (ISS), Bucharest, Romania
\item \Idef{org63}Instituto de Ciencias Nucleares, Universidad Nacional Aut\'{o}noma de M\'{e}xico, Mexico City, Mexico
\item \Idef{org64}Instituto de F\'{\i}sica, Universidad Nacional Aut\'{o}noma de M\'{e}xico, Mexico City, Mexico
\item \Idef{org65}iThemba LABS, National Research Foundation, Somerset West, South Africa
\item \Idef{org66}Joint Institute for Nuclear Research (JINR), Dubna, Russia
\item \Idef{org67}Konkuk University, Seoul, South Korea
\item \Idef{org68}Korea Institute of Science and Technology Information, Daejeon, South Korea
\item \Idef{org69}KTO Karatay University, Konya, Turkey
\item \Idef{org70}Laboratoire de Physique Corpusculaire (LPC), Clermont Universit\'{e}, Universit\'{e} Blaise Pascal, CNRS--IN2P3, Clermont-Ferrand, France
\item \Idef{org71}Laboratoire de Physique Subatomique et de Cosmologie, Universit\'{e} Grenoble-Alpes, CNRS-IN2P3, Grenoble, France
\item \Idef{org72}Laboratori Nazionali di Frascati, INFN, Frascati, Italy
\item \Idef{org73}Laboratori Nazionali di Legnaro, INFN, Legnaro, Italy
\item \Idef{org74}Lawrence Berkeley National Laboratory, Berkeley, California, United States
\item \Idef{org75}Lawrence Livermore National Laboratory, Livermore, California, United States
\item \Idef{org76}Moscow Engineering Physics Institute, Moscow, Russia
\item \Idef{org77}National Centre for Nuclear Studies, Warsaw, Poland
\item \Idef{org78}National Institute for Physics and Nuclear Engineering, Bucharest, Romania
\item \Idef{org79}National Institute of Science Education and Research, Bhubaneswar, India
\item \Idef{org80}Niels Bohr Institute, University of Copenhagen, Copenhagen, Denmark
\item \Idef{org81}Nikhef, Nationaal instituut voor subatomaire fysica, Amsterdam, Netherlands
\item \Idef{org82}Nuclear Physics Group, STFC Daresbury Laboratory, Daresbury, United Kingdom
\item \Idef{org83}Nuclear Physics Institute, Academy of Sciences of the Czech Republic, \v{R}e\v{z} u Prahy, Czech Republic
\item \Idef{org84}Oak Ridge National Laboratory, Oak Ridge, Tennessee, United States
\item \Idef{org85}Petersburg Nuclear Physics Institute, Gatchina, Russia
\item \Idef{org86}Physics Department, Creighton University, Omaha, Nebraska, United States
\item \Idef{org87}Physics Department, Panjab University, Chandigarh, India
\item \Idef{org88}Physics Department, University of Athens, Athens, Greece
\item \Idef{org89}Physics Department, University of Cape Town, Cape Town, South Africa
\item \Idef{org90}Physics Department, University of Jammu, Jammu, India
\item \Idef{org91}Physics Department, University of Rajasthan, Jaipur, India
\item \Idef{org92}Physik Department, Technische Universit\"{a}t M\"{u}nchen, Munich, Germany
\item \Idef{org93}Physikalisches Institut, Ruprecht-Karls-Universit\"{a}t Heidelberg, Heidelberg, Germany
\item \Idef{org94}Politecnico di Torino, Turin, Italy
\item \Idef{org95}Purdue University, West Lafayette, Indiana, United States
\item \Idef{org96}Pusan National University, Pusan, South Korea
\item \Idef{org97}Research Division and ExtreMe Matter Institute EMMI, GSI Helmholtzzentrum f\"ur Schwerionenforschung, Darmstadt, Germany
\item \Idef{org98}Rudjer Bo\v{s}kovi\'{c} Institute, Zagreb, Croatia
\item \Idef{org99}Russian Federal Nuclear Center (VNIIEF), Sarov, Russia
\item \Idef{org100}Russian Research Centre Kurchatov Institute, Moscow, Russia
\item \Idef{org101}Saha Institute of Nuclear Physics, Kolkata, India
\item \Idef{org102}School of Physics and Astronomy, University of Birmingham, Birmingham, United Kingdom
\item \Idef{org103}Secci\'{o}n F\'{\i}sica, Departamento de Ciencias, Pontificia Universidad Cat\'{o}lica del Per\'{u}, Lima, Peru
\item \Idef{org104}Sezione INFN, Bari, Italy
\item \Idef{org105}Sezione INFN, Bologna, Italy
\item \Idef{org106}Sezione INFN, Cagliari, Italy
\item \Idef{org107}Sezione INFN, Catania, Italy
\item \Idef{org108}Sezione INFN, Padova, Italy
\item \Idef{org109}Sezione INFN, Rome, Italy
\item \Idef{org110}Sezione INFN, Trieste, Italy
\item \Idef{org111}Sezione INFN, Turin, Italy
\item \Idef{org112}SSC IHEP of NRC Kurchatov institute, Protvino, Russia
\item \Idef{org113}SUBATECH, Ecole des Mines de Nantes, Universit\'{e} de Nantes, CNRS-IN2P3, Nantes, France
\item \Idef{org114}Suranaree University of Technology, Nakhon Ratchasima, Thailand
\item \Idef{org115}Technical University of Ko\v{s}ice, Ko\v{s}ice, Slovakia
\item \Idef{org116}Technical University of Split FESB, Split, Croatia
\item \Idef{org117}The Henryk Niewodniczanski Institute of Nuclear Physics, Polish Academy of Sciences, Cracow, Poland
\item \Idef{org118}The University of Texas at Austin, Physics Department, Austin, Texas, USA
\item \Idef{org119}Universidad Aut\'{o}noma de Sinaloa, Culiac\'{a}n, Mexico
\item \Idef{org120}Universidade de S\~{a}o Paulo (USP), S\~{a}o Paulo, Brazil
\item \Idef{org121}Universidade Estadual de Campinas (UNICAMP), Campinas, Brazil
\item \Idef{org122}University of Houston, Houston, Texas, United States
\item \Idef{org123}University of Jyv\"{a}skyl\"{a}, Jyv\"{a}skyl\"{a}, Finland
\item \Idef{org124}University of Liverpool, Liverpool, United Kingdom
\item \Idef{org125}University of Tennessee, Knoxville, Tennessee, United States
\item \Idef{org126}University of the Witwatersrand, Johannesburg, South Africa
\item \Idef{org127}University of Tokyo, Tokyo, Japan
\item \Idef{org128}University of Tsukuba, Tsukuba, Japan
\item \Idef{org129}University of Zagreb, Zagreb, Croatia
\item \Idef{org130}Universit\'{e} de Lyon, Universit\'{e} Lyon 1, CNRS/IN2P3, IPN-Lyon, Villeurbanne, France
\item \Idef{org131}V.~Fock Institute for Physics, St. Petersburg State University, St. Petersburg, Russia
\item \Idef{org132}Variable Energy Cyclotron Centre, Kolkata, India
\item \Idef{org133}Vin\v{c}a Institute of Nuclear Sciences, Belgrade, Serbia
\item \Idef{org134}Warsaw University of Technology, Warsaw, Poland
\item \Idef{org135}Wayne State University, Detroit, Michigan, United States
\item \Idef{org136}Wigner Research Centre for Physics, Hungarian Academy of Sciences, Budapest, Hungary
\item \Idef{org137}Yale University, New Haven, Connecticut, United States
\item \Idef{org138}Yonsei University, Seoul, South Korea
\item \Idef{org139}Zentrum f\"{u}r Technologietransfer und Telekommunikation (ZTT), Fachhochschule Worms, Worms, Germany
\end{Authlist}
\endgroup

\end{document}